\documentclass[]{spie}  
\usepackage[]{graphicx}

\usepackage{url}
\usepackage{color, colortbl}
\definecolor{lightgray}{rgb}{0.83, 0.83, 0.83}

\title{The Simons Observatory: Instrument Overview} 

\author{
Nicholas Galitzki\supit{a}, Aamir Ali\supit{b}, Kam S. Arnold\supit{a}, Peter C. Ashton\supit{b,c}, Jason E. Austermann\supit{d}, Carlo Baccigalupi\supit{e}, Taylor Baildon\supit{r}, Darcy Barron\supit{b}, James A. Beall\supit{d}, Shawn Beckman\supit{b}, Sarah Marie M. Bruno\supit{i}, Sean Bryan\supit{f}, Paolo G. Calisse\supit{a,g}, Grace E. Chesmore\supit{r}, Yuji Chinone\supit{b}, Steve K. Choi\supit{i}, Gabriele Coppi\supit{h}, Kevin D. Crowley\supit{a}, Kevin T. Crowley\supit{i}, Ari Cukierman\supit{b}, Mark J. Devlin\supit{h}, Simon Dicker\supit{h}, Bradley Dober\supit{d}, Shannon M. Duff\supit{d}, Jo Dunkley\supit{i}, Giulio Fabbian\supit{j}, Patricio A. Gallardo\supit{k}, Martina Gerbino\supit{l}, Neil Goeckner-Wald\supit{b}, Joseph E. Golec\supit{r}, Jon E. Gudmundsson\supit{l}, Erin E. Healy\supit{i}, Shawn Henderson\supit{y}, Charles A. Hill\supit{b,c}, Gene C. Hilton\supit{d}, Shuay-Pwu Patty Ho\supit{i}, Logan A. Howe\supit{a}, Johannes Hubmayr\supit{d}, Oliver Jeong\supit{b}, Brian Keating\supit{a}, Brian J. Koopman\supit{k}, Kenji Kuichi\supit{m}, Akito Kusaka\supit{c,m}, Jacob Lashner\supit{z}, Adrian T. Lee\supit{b,c,n}, Yaqiong Li\supit{i}, Michele Limon\supit{h}, Marius Lungu\supit{i}, Frederick Matsuda\supit{o}, Philip D. Mauskopf\supit{p}, Andrew J. May\supit{q}, Nialh McCallum\supit{q}, Jeff McMahon\supit{r}, Federico Nati\supit{h}, Michael D. Niemack\supit{k}, John L. Orlowski-Scherer\supit{h}, Stephen C. Parshley\supit{s}, Lucio Piccirillo\supit{t}, Mayuri Sathyanarayana Rao\supit{c,u}, Christopher Raum\supit{b}, Maria Salatino\supit{v}, Joseph S. Seibert\supit{a}, Carlos Sierra\supit{r}, Max Silva-Feaver\supit{a}, Sara M. Simon\supit{r}, Suzanne T. Staggs\supit{i}, Jason R. Stevens\supit{k}, Aritoki Suzuki\supit{c}, Grant Teply\supit{a}, Robert Thornton\supit{w}, Calvin Tsai\supit{a}, Joel N. Ullom\supit{d}, Eve M. Vavagiakis\supit{k}, Michael R. Vissers\supit{d}, Benjamin Westbrook\supit{b}, Edward J. Wollack\supit{x}, Zhilei Xu\supit{h}, Ningfeng Zhu\supit{h} 
\skiplinehalf
\supit{a}Department of Physics, University of California San Diego, La Jolla, CA, USA;\\
\supit{b}Department of Physics, University of California, Berkeley, Berkeley, CA, USA;\\
\supit{c}Physics Division, Lawrence Berkeley National Laboratory, Berkeley, USA;\\
\supit{d}Quantum Sensors Group, NIST, Boulder, CO, USA;\\
\supit{e}SISSA, Trieste, Italy;\\
\supit{f}School of Electrical, Computer and Energy Engineering, Arizona State University, Tempe, AZ, USA;\\
\supit{g}Instituto de Astrof\'isica, Facultad de F\'isica, Pontificia Universidad Cat\'olica de Chile, Macul, Chile;\\
\supit{h}Department of Physics and Astronomy, University of Pennsylvania, Philadelphia, PA, USA;\\
\supit{i}Department of Physics, Princeton University, Princeton, NJ, USA;\\
\supit{j}Institut d`Astrophysique Spatiale, CNRS (UMR 8617), Univ. Paris-Sud, Universit\'e Paris-Saclay, Orsay, France;\\
\supit{k}Department of Physics, Cornell University, Ithaca, NY USA;\\
\supit{l}The Oskar Klein Centre for Cosmoparticle Physics, Department of Physics, Stockholm University, Stockholm, Sweden;\\
\supit{m}Department of Physics, The University of Tokyo, Tokyo, Japan;\\
\supit{n}Radio Astronomy Laboratory, University of California, Berkeley, Berkeley, CA 92093, USA;\\
\supit{o}Kavli IPMU (WPI), The University of Tokyo, Kashiwa, Chiba, Japan;\\
\supit{p}School of Earth and Space Exploration and Department of Physics, Arizona State University, Tempe, AZ, USA;\\
\supit{q}Jodrell Bank Centre for Astrophysics, University of Manchester, Manchester, UK;\\
\supit{r}Department of Physics, University of Michigan, Ann Arbor, USA;\\
\supit{s}Department of Astronomy, Cornell University, Ithaca, NY USA;\\
\supit{t}School of Physics and Astronomy, University of Manchester, UK;\\
\supit{u}Astronomy and Astrophysics, Raman Research Institute, Bangalore, India;\\
\supit{v}AstroParticle and Cosmology laboratory, Paris Diderot University, Paris, France;\\
\supit{w}Department of Physics and Engineering, West Chester University of Pennsylvania, West Chester, PA, USA;\\
\supit{x}NASA/Goddard Space Flight Center, Greenbelt, MD, USA;\\
\supit{y}SLAC National Accelerator Laboratory, Mendlo Park, CA, USA;\\
\supit{z}Department of Physics, University of Southern California, Los Angeles, CA, USA\\
}

\authorinfo{E-mail: ngalitzki@ucsd.edu}

\begin{document} 
  \maketitle 

\begin{abstract}
 The Simons Observatory (SO) will make precise temperature and polarization measurements of the cosmic microwave background (CMB) using a set of telescopes which will cover angular scales between 1 arcminute and tens of degrees, contain over 60,000 detectors, and observe at frequencies between 27 and 270 GHz. SO will consist of a 6\,m aperture telescope coupled to over 30,000 transition-edge sensor bolometers along with three 42\,cm aperture refractive telescopes, coupled to an additional 30,000+ detectors, all of which will be located in the Atacama Desert at an altitude of 5190 m. The powerful combination of large and small apertures in a CMB observatory will allow us to sample a wide range of angular scales over a common survey area. SO will measure fundamental cosmological parameters of our universe, constrain primordial fluctuations, find high redshift clusters via the Sunyaev-Zel’dovich effect, constrain properties of neutrinos, and trace the density and velocity of the matter in the universe over cosmic time. The complex set of technical and science requirements for this experiment has led to innovative instrumentation solutions which we will discuss. The large aperture telescope will couple to a cryogenic receiver that is 2.4\,m in diameter and nearly 3\,m long, creating a number of technical challenges. Concurrently, we are designing the array of cryogenic receivers housing the 42\,cm aperture telescopes. We will discuss the sensor technology SO will use and we will give an overview of the drivers for and designs of the SO telescopes and receivers, with their cold optical components and detector arrays.
 
\end{abstract}




\section{INTRODUCTION}
\label{sec:intro}  

The cosmic microwave background (CMB) has emerged as one of the most powerful probes of the early universe. Measurements of temperature anisotropies on the level of ten parts per million have brought cosmology into a precision era, and have placed tight constraints on the fundamental properties of the Universe\cite{Planck2018}. Beyond temperature anisotropies, CMB polarization anisotropies not only enrich our understanding of our cosmological model, but could potentially provide clues to the very beginning of the universe via the detection (or non-detection) of primordial gravitational waves.  
To provide a complete picture of cosmology, measurements at multiple frequencies of both large and small angular scales are important. 
This is the goal of the Simons Observatory (SO).  

SO will field a 6\,m diameter crossed Dragone\cite{Dragone1978} large aperture telescope (LAT) coupled to the large aperture telescope receiver (LATR) (Sec. \ref{sec:lat}). The LAT is designed to have a large field of view (FOV)\cite{Niemack2016,Parshley2018} (7.2$^\circ$ at 90 GHz) capable of supporting a cryostat with up to 19 tubes, each containing three lenses and a focal plane of detector arrays (henceforth optics tubes). To reduce the development risk of such a large cryostat, the LATR is designed with capacity for 13 optics tubes. During the initial deployment, we plan to populate seven optics tubes with three detector wafers in each for a total of over 30,000 detectors. Note that each optics tube could be upgraded to support four wafers for a $\sim$33\% increase in the number of detectors per optics tube. With this upgrade and the deployment of 19 optics tubes, the LAT focal plane could support roughly 125,000 detectors at 90/150\,GHz. 

SO will also deploy an array of 42 cm aperture small aperture telescopes (SATs) with seven detector wafers in each coupled to over 30,000 detectors (Sec. \ref{sec:sat}). SO will observe with three frequency pairs in order to observe the CMB peak signal and constrain polarized foreground contamination from galactic synchotron and dust emission at lower and higher frequencies. The SO frequency bands are: 27/39 GHz, low frequency(LF); 90/150 GHz, mid-frequency(MF); and 220/270 GHz, ultra-high frequency(UHF). The LATR will initially be populated with four MF optics tubes, two UHF optics tubes, and one LF optics tube, while the SAT array will initially be composed of two MF SATs and one UHF SAT with a fourth LF SAT to follow.
Details of the SO observing strategy with the two classes of telescope can be found in Stevens et al. 2018\cite{Stevens2018}. A description of the sensitivity calculator used to optimize the SO instrument design appears in Hill et al. 2018\cite{Hill2018}. Details of the calibration strategy planned for SO can be found in Bryan et al. 2018\cite{Bryan2018}. Additional details of the forecasting studies which led to the SO instrument configuration as well as a more in depth description of the observatory will be in a pair of papers in preparation. 


In this paper we will introduce the basic design of the detector systems, telescopes, and receivers, many of which are discussed in more detail in a set of complementary papers referenced herein. Sec. \ref{sec:sat} describes the SAT in additional detail as this paper serves as the most comprehensive reference for the SAT instrument description.

\section{Sensor technology}
\label{sec:sensors}

   \begin{figure}[t]
    	\begin{center}
        \includegraphics[width = 1.0\linewidth]{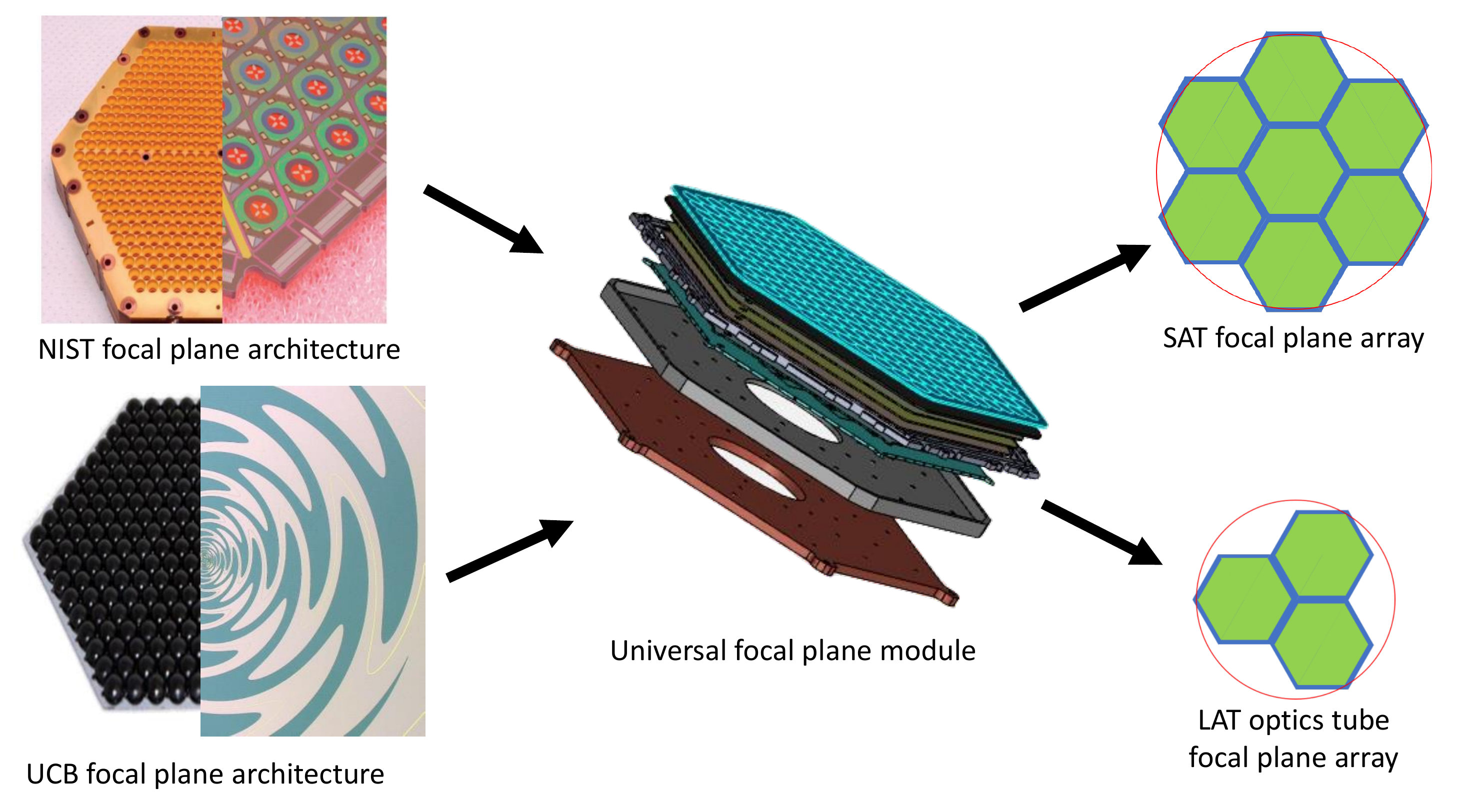}
        \end{center}
    	\caption[Caption for UFM]{Left: Illustrative images of the two detector architectures with the optical coupling structure at left and the planar structure of the detector wafers on the right. The top left image shows a feedhorn array which is coupled to an array of OMTs from NIST. The bottom left image shows a lenslet array which couples to sinuous antennas from UCB. Each type will be mounted into a universal focal plane module (UFM) shown at center which couples the detector array to the cold readout wafers, magnetic shielding, and mechanical coupling. The UFMs are then assembled into focal planes shown at right with seven UFMs in a SAT focal plane and three UFMs per LAT optics tube.} 
        \label{fig:ufm}
    \end{figure}

SO uses AlMn transition-edge sensor (TES)\cite{Li2016} bolometers fabricated on 150\,mm diameter silicon wafers with two demonstrated technologies for radio frequency (RF) coupling (Fig. \ref{fig:ufm}): lenslet coupled sinuous antennas\cite{Suzuki2012} and horn coupled orthomode transducers (OMTs)\cite{Mcmahon2012, Choi2018, Simon2016, Henderson2016}.  The bolometer arrays with sinuous antennas will be fabricated at the University of California - Berkeley (UCB) while the ones with OMTs will be fabricated at the National Institute of Standards and Technology (NIST)\cite{Duff2016}. The TESs are read out with microwave SQUID multiplexing ($\mu$mux) \cite{Mates2011, Dober2017} which has been demonstrated in the MUSTANG2 camera\cite{Dicker2014}. The focal plane and readout architecture were chosen to optimize sensitivity with a balance of technology development, cost risk, and schedule risk. Additional details on the feedhorn development and lenslet development for SO are presented in Simon et al. 2018\cite{Simon2018} and Beckman et al. 2018\cite{Beckman2018}, respectively. Each detector pixel has four bolometers to sense two orthogonal polarizations in each of the two frequency bands. The TESs are tuned to have a $T_c\approx160$\,mK. More details of the systematics associated with both detector array types and a discussion of readout associated crosstalk can be found in Crowley et al. 2018\cite{Crowley2018}. 


\subsection{Microwave multiplexing readout electronics}


The $\mu$mux readout technology is capable of reading out thousands of detectors on a single pair of RF lines. The high multiplexing factor vastly simplifies the cabling for the detectors which must penetrate a vacuum shell and multiple radiation shields. Additional details of the $\mu$mux system can be found in Dober et al. 2017\cite{Dober2017} and Henderson et al. 2018\cite{Henderson2018}.


The detector wafer and readout components are stacked together in a universal focal plane module (UFM) which has RF connectors for the RF input and output signals as well as a single connector for flux modulation and TES bias lines. The detector biasing and cryogenic readout components comprise several dozen silicon chips and a stack of three silicon wafers which mount behind the detector wafer as shown in the middle image of Fig.~\ref{fig:ufm}. UFMs will be tested as independent units prior to installation into three wafer or seven wafer focal plane arrays (FPAs) for the LATR or SAT, respectively as shown in Fig. \ref{fig:ufm}. More details on the UFM assembly can be found in Ho et al. 2018\cite{Ho2018}.

\section{Large aperture telescope}
\label{sec:lat}
   \begin{figure}[t]
    	\begin{center}
        \begin{tabular}{c}
        \includegraphics[width = 1.0\linewidth]{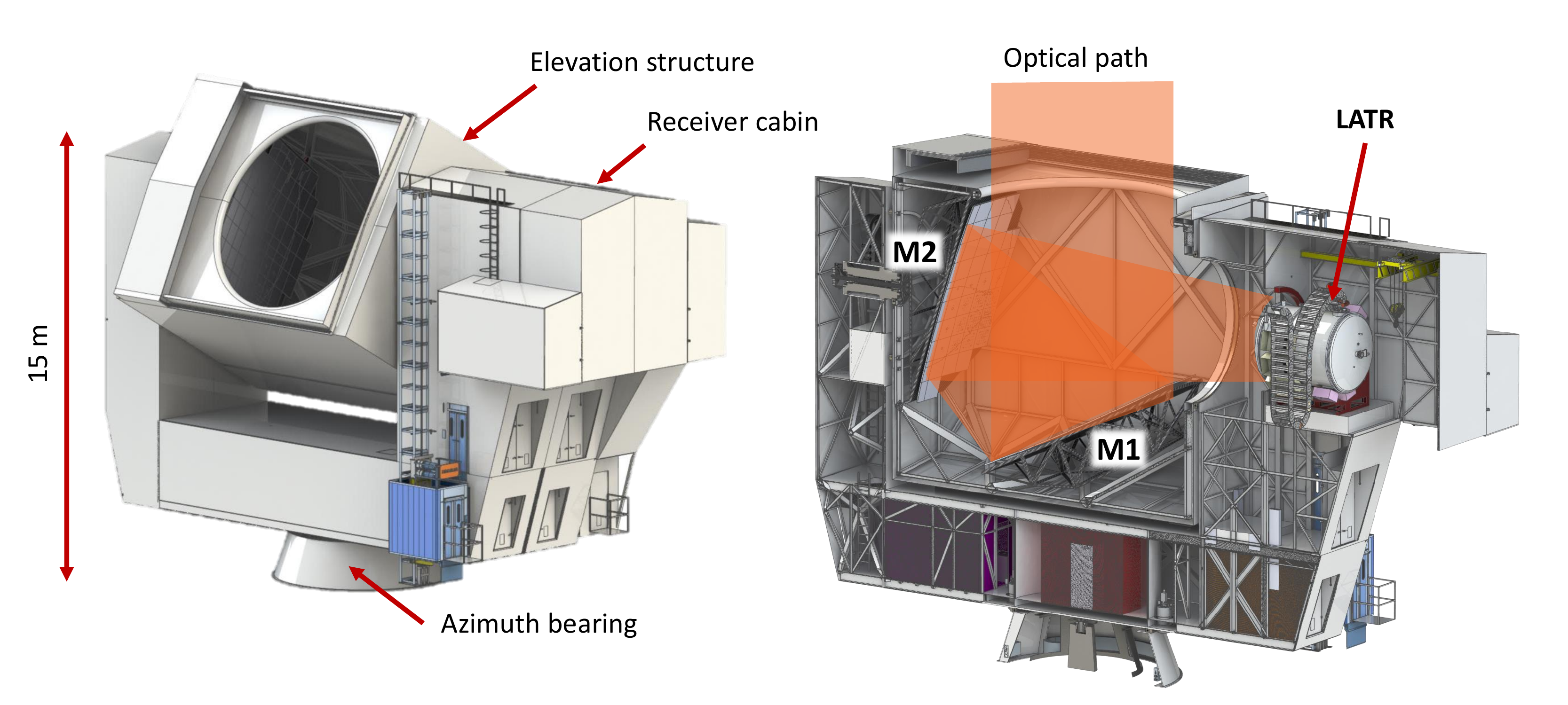}
        \end{tabular}
        \end{center}
    	\caption{A cross-section showing how the LATR couples to the LAT.  In the orientation shown on the right, light enters the telescope from the top.  It then reflects off the 6\,m primary (M1) and 6\,m secondary (M2) before being directed to the receiver. The receiver always operates in the horizontal orientation shown here; it is not directly mechanically coupled to the telescope elevation structure.  Therefore, a separate mechanism is used to rotate the receiver about its long axis as the telescope elevation structure moves M1 and M2 in rotation.
        } 
        \label{fig:telescope}
    \end{figure}

A crossed Dragone design concept was chosen for the SO LAT. The LAT design is nearly identical to that of the Cerro Chajnantor Atacama Telescope prime (CCATp)\cite{Parshley2018_2}. CCATp allowed SO to build off their existing development, reducing the schedule risk for delivery of the LAT. Both the SO LAT and CCATp will be built by Vertex Antennentechnik GmbH\footnote{Vertex Antennentechnik GmbH, Duisburg, Germany} with the LAT delivery to the SO site scheduled for 2020. Details of the design can be found in Parshley et al. 2018\cite{Parshley2018} and an overview is shown in Fig. \ref{fig:telescope}. The refractive reimaging optics for each optics tube are composed of three silicon lenses with machined metamaterial antireflection (AR) surface layers\cite{Datta2013}. Details of the optimization of the LAT optical design appear in Dicker et al. 2018\cite{Dicker2018}.


The optical design was optimized using ZEMAX and a full analysis of systematics effects carried out using ZEMAX and GRASP.  Some of the effects studied encompassed tolerancing, pointing errors, optical distortions, beam ellipticity, cross polar response, instrumental polarization rotation, and sidelobe pickup. 
Additional details on the systematics studies of the LAT optical design can be found in Gallardo et al. 2018\cite{Gallardo2018}.

\subsection{Large aperture telescope receiver}

\begin{figure}[t]
  \centerline{
    \includegraphics[height=3.0in]{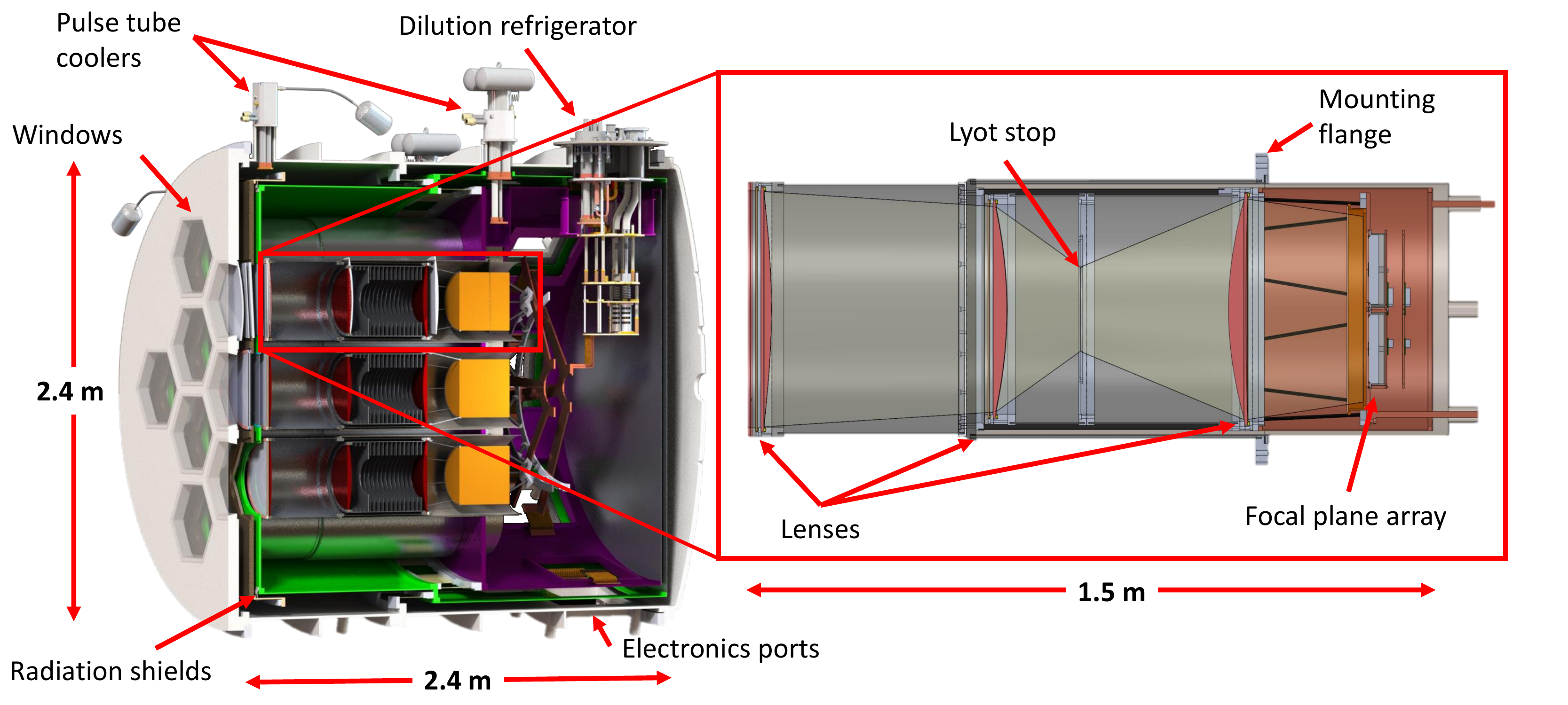}
  }
  \caption{Cross sectional view of the LATR with primary components labeled. A cross section of a single optics tube is detailed at right with the main optical components labeled.
    \label{fig:latr_1}}
\end{figure}

The LATR is designed around 13 tightly spaced optics tubes. The optics tubes couple to a 4\,K stage which cools the first lens and optics tube body. A 1\,K stage cools the second and third lens and Lyot stop, with the detector array coupled to a 100\,mK stage. The optics tubes are modular and capable of being inserted from the back of the cryostat while it is mounted on the telescope. 
The LATR has stages cooled to 80\,K, 40\,K, and 4\,K by pulse tube coolers and 1\,K and 100\,mK stages cooled by a Bluefors dilution refrigerator\footnote{Bluefors Cryogenics, Helsinki, Finland}. An overview of the receiver and optics tube design is shown in Fig. \ref{fig:latr_1}. More details about the mechanical and cryogenic design of the receiver can be found in Zhu et al. 2018\cite{Zhu2018} and Orlowski-Scherer et al. 2018\cite{Orlowski-Scherer2018}. Simulations were also run on the cool down process for the LATR which predict it will take up to a month to reach base temperatures without an effective heat switch strategy as described in Coppi et al. 2018\cite{Coppi2018}. Note that the first light camera for CCATp will share many elements of the LATR design. Details of the CCATp seven optics tube receiver can be found in Vavagiakis et al. 2018\cite{Vavagiakis2018}.

\section{Small aperture telescope}
\label{sec:sat}

   \begin{figure}[t]
    	\begin{center}
        \includegraphics[width = 1.0\linewidth]{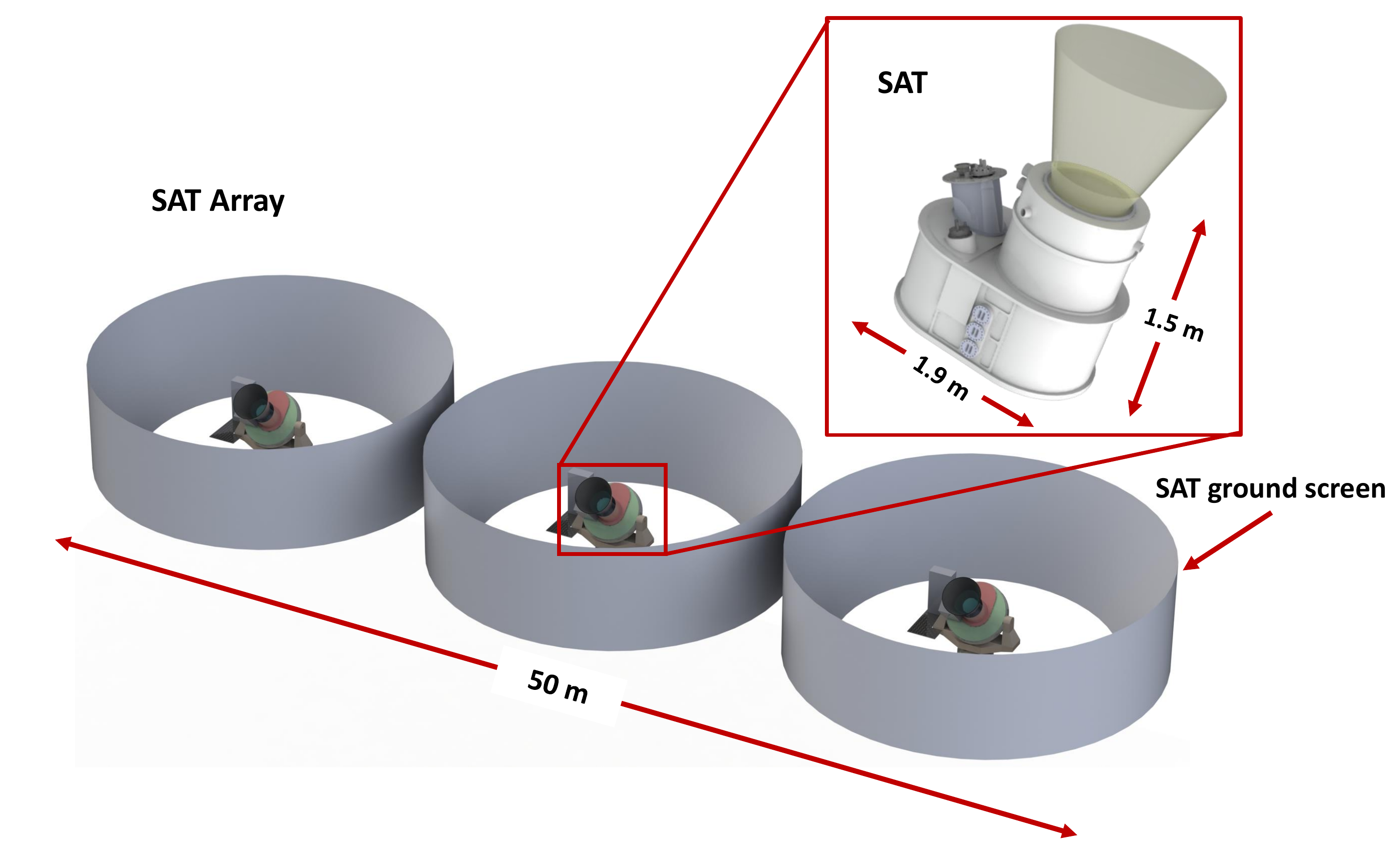}
        \end{center}
    	\caption[Caption for SAT Array]{A conceptual image of the SAT array consisting of three telescopes on pointing platforms with each surrounded by a ground shield. The inset shows a single SAT receiver with the rays from the 35$^\circ$ FOV beam shown as the shaded cone.
} 
        \label{fig:satarr}
    \end{figure}
The SO SAT is designed to observe CMB polarization signals on degree angular scales, where a faint peak in the parity-odd polarization signal, over four orders of magnitude smaller than CMB temperature anisotropies, is predicted to occur. The SAT science goals require a deep survey area with impeccable systematics control.
Optimizing the throughput of the SAT provided the simplest path to maximizing mapping speed which resulted in a short focal length system with a large FOV (35$^\circ$) that couples to seven 150\,mm detector wafers. The SAT design increases mapping speed by cooling the aperture stop and lenses to 1\,K. The SAT receiver will also house a continuously rotating cryogenic half-wave plate (CHWP) between the receiver window and first lens to modulate the polarized signal to allow for additional systematics control. 

\subsection{Optical design}

  The large focal plane area combined with restrictions from the CHWP and lenses on the aperture stop size formed the principal constraints on the optical design. The location of the CHWP in front of the aperture stop (see Sec. \ref{sec:chwp}) effectively determined the stop size as the sapphire used to make the CHWP has a maximum diameter of 55\,cm. Given that the stop is coupled to the 1\,K thermal stage while the CHWP is mounted to the warmer 40\,K thermal stage, 42\,cm in diameter was determined to be the largest stop diameter possible, taking spacing and mounting requirements into consideration. The thermal and optical constraints on the system effectively narrowed down the possible optical designs for the SAT.

We decided to use a design with three silicon lenses, with metamaterial AR surfaces as our primary option. However, the cryogenic volume is sized to accommodate a hybrid three-lens design which achieves slightly better performance by combining two silicon lenses with a 65\,cm diameter alumina lens. A direct benefit of intentionally increasing the size of the optics volume in diameter was the ability to place radially deep 1\,K baffles along the entire tube length which will effectively control the sidelobe response. 
Additional performance optimization of the optical design was done to include the effects of filters, especially the alumina filters, and the ultra high molecular weight (UHMWPE) window. The left panel of Fig. \ref{fig:sat} shows the position of the lenses within a MF SAT.

\subsection{Platform}
\label{sec:platform}

The SAT platform is designed to allow rotation around the boresight axis for additional systematics controls. The operational tilt range of $<40^\circ$ from vertical of a pulse tube cooler (PTC) allows a boresight rotation of +/- 75$^\circ$ at a pointing angle higher than 50$^\circ$ above the horizon. The constraints on boresight rotation range combined with azimuth and elevation pointing requirements drive the pointing platform design and manufacture, which is part of future work not included in this paper.

\subsection{SAT cryogenic receiver}

\begin{figure}[t]
  \centerline{
    \includegraphics[height=3.0in]{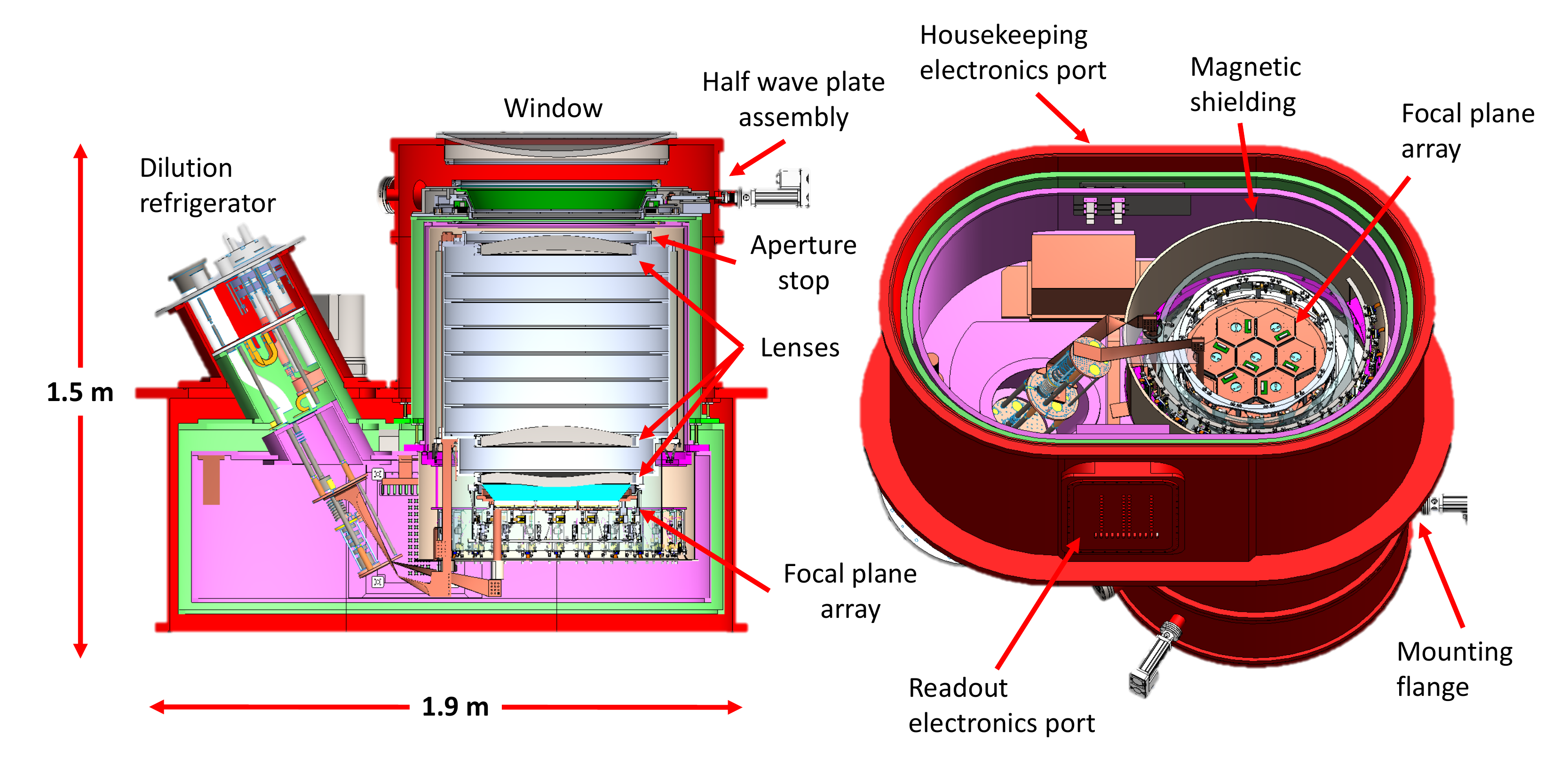}
  }
  \caption{Detailed views of the primary components of the SAT receiver. The image at left shows a cross-sectional view highlighting the optics tube and focal plane array placement as well as the angled dilution refrigerator. The view at right shows the back end of the receiver with the lid section cut away showing the primary components as they are situated in the 4\,K volume. The PT420 unit used to cool the 40\,K and 4\,K stages is adjacent to the dilution refrigerator and is not visible in these images.
    \label{fig:sat}}
\end{figure}

The optical design and cryogenic tilt angle determined the overall structure of the SAT receiver. The SAT receiver is composed of a primary cylindrical volume surrounding the optics tube and focal plane combined with a secondary volume to accommodate the cryogenic systems as shown in Fig. \ref{fig:sat}. The CHWP requires an additional space in front of the optics tube for mounting and operation. The readout systems for the detectors also require a dedicated electronics port in the side of the cryostat that presented a significant design challenge to ensure appropriate thermal coupling, cable routing, and component mounting for the readout chain operational requirements.

The design temperature of each stage was determined by the detector and optical requirements. The SAT uses a Cryomech PT420 to cool a 40\,K stage and a 4\,K stage and a Bluefors SD400 to cool a 1\,K and 100\,mK stage. The 40\,K stage is used primarily as a radiation shield, an intercept stage for conductive elements, and to remove optical IR loading through the window. Additionally, the 40\,K stage couples to the CHWP which requires an operational temperature below its superconducting transition temperature of 90\,K 	(see section \ref{sec:chwp}). The 4\,K volume is an additional shield layer that surrounds the components providing the thermal distribution from the DR's 1\,K and 0.1\,K stages. The 1\,K stage cools the aperture stop and lenses as well as providing an additional intercept stage to conductive elements that go to 100\,mK. The 100\,mK stage cools the FPA. 

A primary design goal for the receiver was to keep the overarching mechanical design concept as simple as possible which we accomplished with a set of mechanical trusses coupling different temperature stages near their respective centers of mass along the vertical axis. A two layer G10 tab Vierendeel truss\cite{Wickersheimer1976} connects the vacuum shell to the 40\,K and 4\,K stages providing a rigid coupling that is designed to accommodate the differential thermal contraction between the layers. The truss builds on the experience of the POLARBEAR-2 truss\cite{Inoue2016} combined with the successful tab structure implemented in SPIDER\cite{Gudmundsson2015}. Thermally isolating mechanical connections between the 4\,K stage and 1\,K optics tube and between the 1\,K optics tube and the 100\,mK focal plane are provided by two additional truss structures that utilize pultruded carbon fiber tubes to provide mechanical support and thermal isolation. 

\subsubsection{Thermal model and cooling configuration}
\label{sec:thermal}

\begin{table}
\vspace{3mm}
\begin{centering}
\setlength\arrayrulewidth{0.75pt}
\begin{tabular}[t]{|c|c|c|c|c|c|c|}
	\hline
	\multicolumn{7}{|l|}{{\bf SAT Thermal Loading Estimates}}\\
	\hline
	 \rowcolor{lightgray} Stage & Support & Cabling & Radiative & Optical & \bf{Total} & \bf{Available} \\
	\hline
    	40\,K (PTC) & 5.2\,W & 4.5\,W & 6.5\,W & 6.8\,W & 23\,W & 55\,W\\
     \hline
     4\,K (PTC) & 180\,mW & 200\,mW & 4\,mW & 36\,mW & 420\,mW & 2000\,mW\\
     \hline
     1\,K (DR) & 1\,mW & 1\,mW & 1\,mW & 5\,mW & 8\,mW & 20\,mW\\
     \hline
     0.1\,K (DR) & 3\,$\mu$W & 10\,$\mu$W & 1\,$\mu$W & 1\,$\mu$W & 15\,$\mu$W & 400\,$\mu$W\\
     \hline
     \end{tabular}
     \vspace{3mm}
     \caption{Loading estimates for each temperature stage of the receiver split by source. `Support' encapsulates structural elements while `optical' refers to the output of python simulation code that estimates the loading from filtering elements illuminated by the window aperture. The available' column refers to the total cooling power provided by the respective pulse tube cooler and dilution refrigerator stages. The SAT is designed to have margin at all cooling stages to allow for loading sources not considered in the model either by omission or simplification.}
     \label{tab:loading}
     \end{centering}
\end{table}

The SAT cooling system design necessitated a detailed thermal model that could account for all sources of thermal loading on each of the stages. The majority of the model is encapsulated in an excel spreadsheet capable of performing calculations for all loading types except the loading through the optical path, which was estimated using a separate python script.

The total thermal loading estimates for each stage dictated our overall cooling strategy. Previous experiments have found estimates using the described methods are typically lower than observed as they do not account for many of the imperfections inherent to a physical cryogenic system, including necessary gaps in the multi layer insulation (MLI) blankets, limited information on cryogenic material properties, thermal gradients across stages, and non-ideal filter performance. We have designed the cryogenic system to have twice as much cooling power available compared to estimated loading power to provide sufficient overhead. The predicted loading for the SAT is shown in Table \ref{tab:loading} along with the anticipated cooling power provided by the PT420 and the SD400 units.

\subsubsection{Cryogenic half-wave plate}
\label{sec:chwp}

The SAT includes a continuously rotating sapphire CHWP\cite{Hill2018_2} which modulates the polarization signal from the sky in order to control for a number of systematics\cite{Kusaka2018}. The CHWP is cooled by the first cryogenic stage to minimize the loading on the system. A study of systematics associated with the SO CHWP itself is presented in Salatino et al. 2018\cite{Salatino2018}. The CHWP is placed on the nominal 40\,K stage as close to the aperture as possible to better minimize the amount of instrumental polarization leakage that is modulated into the science band. The only receiver components between the CHWP and the sky are the UHMWPE window, two thin film metal-mesh IR blocking filters\cite{Ade2006}, and the first alumina filter. 


\section{Conclusion}

We have presented an overview of the principal components of SO as well as their current status and the design choices that led to our final instrument configuration. SO is on target to begin scientific observations of the CMB with the LAT and SAT starting in the year 2021. Future publications will detail the performance of the various components as they are built and tested. SO will be one of the most sensitive broadband, wide angular scale, CMB survey instruments to date. With its initial deployment of over 60,000 detectors, SO will provide an important step forward in CMB science and pave the way for future millimeter wave experiments such as CMB-S4\cite{S4sci, S4tech}.

\appendix    

\acknowledgments     
This work was supported in part by a grant from the Simons Foundation (Award \#457687, B.K.)


\bibliography{arxiv_bib}   

\begin{thebibliography}{10}

\bibitem{Planck2018}
{Planck Collaboration}, {Akrami}, Y., {Arroja}, F., {Ashdown}, M., {Aumont},
  J., {Baccigalupi}, C., {Ballardini}, M., {Banday}, A.~J., {Barreiro}, R.~B.,
  {Bartolo}, N., {Basak}, S., {Battye}, R., {Benabed}, K., {Bernard}, J.-P.,
  {Bersanelli}, M., {Bielewicz}, P., {Bock}, J.~J., {Bond}, J.~R., {Borrill},
  J., {Bouchet}, F.~R., {Boulanger}, F., {Bucher}, M., {Burigana}, C.,
  {Butler}, R.~C., {Calabrese}, E., {Cardoso}, J.-F., {Carron}, J.,
  {Casaponsa}, B., {Challinor}, A., {Chiang}, H.~C., {Colombo}, L.~P.~L.,
  {Combet}, C., {Contreras}, D., {Crill}, B.~P., {Cuttaia}, F., {de Bernardis},
  P., {de Zotti}, G., {Delabrouille}, J., {Delouis}, J.-M., {D{\'e}sert},
  F.-X., {Di Valentino}, E., {Dickinson}, C., {Diego}, J.~M., {Donzelli}, S.,
  {Dor{\'e}}, O., {Douspis}, M., {Ducout}, A., {Dupac}, X., {Efstathiou}, G.,
  {Elsner}, F., {En{\ss}lin}, T.~A., {Eriksen}, H.~K., {Falgarone}, E.,
  {Fantaye}, Y., {Fergusson}, J., {Fernandez-Cobos}, R., {Finelli}, F.,
  {Forastieri}, F., {Frailis}, M., {Franceschi}, E., {Frolov}, A., {Galeotta},
  S., {Galli}, S., {Ganga}, K., {G{\'e}nova-Santos}, R.~T., {Gerbino}, M.,
  {Ghosh}, T., {Gonz{\'a}lez-Nuevo}, J., {G{\'o}rski}, K.~M., {Gratton}, S.,
  {Gruppuso}, A., {Gudmundsson}, J.~E., {Hamann}, J., {Handley}, W., {Hansen},
  F.~K., {Helou}, G., {Herranz}, D., {Hivon}, E., {Huang}, Z., {Jaffe}, A.~H.,
  {Jones}, W.~C., {Karakci}, A., {Keih{\"a}nen}, E., {Keskitalo}, R.,
  {Kiiveri}, K., {Kim}, J., {Kisner}, T.~S., {Knox}, L., {Krachmalnicoff}, N.,
  {Kunz}, M., {Kurki-Suonio}, H., {Lagache}, G., {Lamarre}, J.-M., {Langer},
  M., {Lasenby}, A., {Lattanzi}, M., {Lawrence}, C.~R., {Le Jeune}, M.,
  {Leahy}, J.~P., {Lesgourgues}, J., {Levrier}, F., {Lewis}, A., {Liguori}, M.,
  {Lilje}, P.~B., {Lilley}, M., {Lindholm}, V., {L{\'o}pez-Caniego}, M.,
  {Lubin}, P.~M., {Ma}, Y.-Z., {Mac{\'{\i}}as-P{\'e}rez}, J.~F., {Maggio}, G.,
  {Maino}, D., {Mandolesi}, N., {Mangilli}, A., {Marcos-Caballero}, A.,
  {Maris}, M., {Martin}, P.~G., {Mart{\'{\i}}nez-Gonz{\'a}lez}, E.,
  {Matarrese}, S., {Mauri}, N., {McEwen}, J.~D., {Meerburg}, P.~D., {Meinhold},
  P.~R., {Melchiorri}, A., {Mennella}, A., {Migliaccio}, M., {Millea}, M.,
  {Mitra}, S., {Miville-Desch{\^e}nes}, M.-A., {Molinari}, D., {Moneti}, A.,
  {Montier}, L., {Morgante}, G., {Moss}, A., {Mottet}, S., {M{\"u}nchmeyer},
  M., {Natoli}, P., {N{\o}rgaard-Nielsen}, H.~U., {Oxborrow}, C.~A., {Pagano},
  L., {Paoletti}, D., {Partridge}, B., {Patanchon}, G., {Pearson}, T.~J.,
  {Peel}, M., {Peiris}, H.~V., {Perrotta}, F., {Pettorino}, V., {Piacentini},
  F., {Polastri}, L., {Polenta}, G., {Puget}, J.-L., {Rachen}, J.~P.,
  {Reinecke}, M., {Remazeilles}, M., {Renzi}, A., {Rocha}, G., {Rosset}, C.,
  {Roudier}, G., {Rubi{\~n}o-Mart{\'{\i}}n}, J.~A., {Ruiz-Granados}, B.,
  {Salvati}, L., {Sandri}, M., {Savelainen}, M., {Scott}, D., {Shellard},
  E.~P.~S., {Shiraishi}, M., {Sirignano}, C., {Sirri}, G., {Spencer}, L.~D.,
  {Sunyaev}, R., {Suur-Uski}, A.-S., {Tauber}, J.~A., {Tavagnacco}, D.,
  {Tenti}, M., {Terenzi}, L., {Toffolatti}, L., {Tomasi}, M., {Trombetti}, T.,
  {Valiviita}, J., {Van Tent}, B., {Vibert}, L., {Vielva}, P., {Villa}, F.,
  {Vittorio}, N., {Wandelt}, B.~D., {Wehus}, I.~K., {White}, M., {White},
  S.~D.~M., {Zacchei}, A., and {Zonca}, A., ``{Planck 2018 results. I. Overview
  and the cosmological legacy of Planck},'' {\em ArXiv e-prints}  (July 2018).

\bibitem{Dragone1978}
{Dragone}, C., ``{Offset multireflector antennas with perfect pattern symmetry
  and polarization discrimination},'' {\em AT T Technical Journal}~{\bf 57},
  2663--2684 (Sept. 1978).

\bibitem{Niemack2016}
Niemack, M.~D., ``{Designs for a large-aperture telescope to map the CMB 10
  faster},'' {\em \tt arXiv:1511.04506v2 [astro-ph.IM]}  (Feb 2018).

\bibitem{Parshley2018}
Parshley, S.~C. et~al., ``The optical design of the six-meter {CCAT-prime and
  Simons Observatory telescopes},'' in [{\em Ground-based and Airborne
  Telescopes VII}{\nolinebreak\hspace{0.1em}]},  {\em Proc. SPIE} ,  10700 --
  145 (2018).

\bibitem{Stevens2018}
Stevens, J. et~al., ``{Designs for Next Generation CMB Survey Strategies from
  Chile},'' in [{\em Millimeter, Submillimeter, and Far-Infrared Detectors and
  Instrumentation for Astronomy IX}{\nolinebreak\hspace{0.1em}]},  {\em
  \procspie},  10708--136 (2018).

\bibitem{Hill2018}
Hill, C.~A., Bruno, S. M.~M., Simon, S.~M., Ali, A., Arnold, K.~S., Ashton,
  P.~C., Barron, D., Bryan, S., Chinone, Y., Coppi, G., Crowley, K.~T.,
  Cukierman, A., Dicker, S., Dunkley, J., Fabbian, G., Galitzki, N., Gallardo,
  P.~A., Gudmundsson, J.~E., Hubmayr, J., Keating, B., Kusaka, A., Lee, A.~T.,
  Matsuda, F., Mauskopf, P.~D., McMahon, J., Niemack, M.~D., Puglisi, G., Rao,
  M.~S., Salatino, M., Sierra, C., Staggs, S., Suzuki, A., Teply, G., Ullom,
  J.~N., Westbrook, B., Xu, Z., and Zhu, N., ``Bolocalc: a sensitivity
  calculator for the design of simons observatory,'' {\em Proc.SPIE}~{\bf
  10708},  10708 -- 10708 -- 21 (2018).

\bibitem{Bryan2018}
Bryan, S.~A., Simon, S.~M., Gerbino, M., Teply, G., Ali, A., Chinone, Y.,
  Crowley, K., Fabbian, G., Gallardo, P.~A., Goeckner-Wald, N., Keating, B.,
  Koopman, B., Kusaka, A., Matsuda, F., Mauskopf, P., McMahon, J., Nati, F.,
  Puglisi, G., Reichardt, C.~L., Salatino, M., Xu, Z., and Zhu, N.,
  ``Development of calibration strategies for the simons observatory,'' {\em
  Proc.SPIE}~{\bf 10708},  10708 -- 10708 -- 13 (2018).

\bibitem{Li2016}
{Li}, D., {Austermann}, J.~E., {Beall}, J.~A., {Becker}, D.~T., {Duff}, S.~M.,
  {Gallardo}, P.~A., {Henderson}, S.~W., {Hilton}, G.~C., {Ho}, S.-P.,
  {Hubmayr}, J., {Koopman}, B.~J., {McMahon}, J.~J., {Nati}, F., {Niemack},
  M.~D., {Pappas}, C.~G., {Salatino}, M., {Schmitt}, B.~L., {Simon}, S.~M.,
  {Staggs}, S.~T., {Van Lanen}, J., {Ward}, J.~T., and {Wollack}, E.~J.,
  ``{AlMn Transition Edge Sensors for Advanced ACTPol},'' {\em Journal of Low
  Temperature Physics}~{\bf 184},  66--73 (July 2016).

\bibitem{Suzuki2012}
{Suzuki}, A., {Arnold}, K., {Edwards}, J., {Engargiola}, G., {Ghribi}, A.,
  {Holzapfel}, W., {Lee}, A., {Meng}, X., {Myers}, M., {O'Brient}, R.,
  {Quealy}, E., {Rebeiz}, G., and {Richards}, P., ``{Multi-chroic
  Dual-Polarization Bolometric Focal Plane for Studies of the Cosmic Microwave
  Background},'' {\em Journal of Low Temperature Physics}~{\bf 167},  852--858
  (June 2012).

\bibitem{Mcmahon2012}
{McMahon}, J., {Beall}, J., {Becker}, D., {Cho}, H.~M., {Datta}, R., {Fox}, A.,
  {Halverson}, N., {Hubmayr}, J., {Irwin}, K., {Nibarger}, J., {Niemack}, M.,
  and {Smith}, H., ``{Multi-chroic Feed-Horn Coupled TES Polarimeters},'' {\em
  Journal of Low Temperature Physics}~{\bf 167},  879--884 (June 2012).

\bibitem{Choi2018}
{Choi}, S.~K., {Austermann}, J., {Beall}, J.~A., {Crowley}, K.~T., {Datta}, R.,
  {Duff}, S.~M., {Gallardo}, P.~A., {Ho}, S.~P., {Hubmayr}, J., {Koopman},
  B.~J., {Li}, Y., {Nati}, F., {Niemack}, M.~D., {Page}, L.~A., {Salatino}, M.,
  {Simon}, S.~M., {Staggs}, S.~T., {Stevens}, J., {Ullom}, J., and {Wollack},
  E.~J., ``{Characterization of the Mid-Frequency Arrays for Advanced
  ACTPol},'' {\em Journal of Low Temperature Physics}  (June 2018).

\bibitem{Simon2016}
{Simon}, S.~M., {Austermann}, J., {Beall}, J.~A., {Choi}, S.~K., {Coughlin},
  K.~P., {Duff}, S.~M., {Gallardo}, P.~A., {Henderson}, S.~W., {Hills}, F.~B.,
  {Ho}, S.-P.~P., {Hubmayr}, J., {Josaitis}, A., {Koopman}, B.~J., {McMahon},
  J.~J., {Nati}, F., {Newburgh}, L., {Niemack}, M.~D., {Salatino}, M.,
  {Schillaci}, A., {Schmitt}, B.~L., {Staggs}, S.~T., {Vavagiakis}, E.~M.,
  {Ward}, J., and {Wollack}, E.~J., ``{The design and characterization of
  wideband spline-profiled feedhorns for Advanced ACTPol},'' in [{\em
  Millimeter, Submillimeter, and Far-Infrared Detectors and Instrumentation for
  Astronomy VIII}{\nolinebreak\hspace{0.1em}]},  {\em \procspie} {\bf 9914},
  991416 (July 2016).

\bibitem{Henderson2016}
{Henderson}, S.~W., {Allison}, R., {Austermann}, J., {Baildon}, T.,
  {Battaglia}, N., {Beall}, J.~A., {Becker}, D., {De Bernardis}, F., {Bond},
  J.~R., {Calabrese}, E., {Choi}, S.~K., {Coughlin}, K.~P., {Crowley}, K.~T.,
  {Datta}, R., {Devlin}, M.~J., {Duff}, S.~M., {Dunkley}, J., {D{\"u}nner}, R.,
  {van Engelen}, A., {Gallardo}, P.~A., {Grace}, E., {Hasselfield}, M.,
  {Hills}, F., {Hilton}, G.~C., {Hincks}, A.~D., {Hlo{\^z}ek}, R., {Ho}, S.~P.,
  {Hubmayr}, J., {Huffenberger}, K., {Hughes}, J.~P., {Irwin}, K.~D.,
  {Koopman}, B.~J., {Kosowsky}, A.~B., {Li}, D., {McMahon}, J., {Munson}, C.,
  {Nati}, F., {Newburgh}, L., {Niemack}, M.~D., {Niraula}, P., {Page}, L.~A.,
  {Pappas}, C.~G., {Salatino}, M., {Schillaci}, A., {Schmitt}, B.~L., {Sehgal},
  N., {Sherwin}, B.~D., {Sievers}, J.~L., {Simon}, S.~M., {Spergel}, D.~N.,
  {Staggs}, S.~T., {Stevens}, J.~R., {Thornton}, R., {Van Lanen}, J.,
  {Vavagiakis}, E.~M., {Ward}, J.~T., and {Wollack}, E.~J., ``{Advanced ACTPol
  Cryogenic Detector Arrays and Readout},'' {\em Journal of Low Temperature
  Physics}~{\bf 184},  772--779 (Aug. 2016).

\bibitem{Duff2016}
{Duff}, S.~M., {Austermann}, J., {Beall}, J.~A., {Becker}, D., {Datta}, R.,
  {Gallardo}, P.~A., {Henderson}, S.~W., {Hilton}, G.~C., {Ho}, S.~P.,
  {Hubmayr}, J., {Koopman}, B.~J., {Li}, D., {McMahon}, J., {Nati}, F.,
  {Niemack}, M.~D., {Pappas}, C.~G., {Salatino}, M., {Schmitt}, B.~L., {Simon},
  S.~M., {Staggs}, S.~T., {Stevens}, J.~R., {Van Lanen}, J., {Vavagiakis},
  E.~M., {Ward}, J.~T., and {Wollack}, E.~J., ``{Advanced ACTPol Multichroic
  Polarimeter Array Fabrication Process for 150 mm Wafers},'' {\em Journal of
  Low Temperature Physics}~{\bf 184},  634--641 (Aug. 2016).

\bibitem{Mates2011}
{Mates}, J.~A.~B., {\em {The Microwave SQUID Multiplexer}}, PhD thesis,
  University of Colorado at Boulder (2011).

\bibitem{Dober2017}
{Dober}, B., {Becker}, D.~T., {Bennett}, D.~A., {Bryan}, S.~A., {Duff}, S.~M.,
  {Gard}, J.~D., {Hays-Wehle}, J.~P., {Hilton}, G.~C., {Hubmayr}, J., {Mates},
  J.~A.~B., {Reintsema}, C.~D., {Vale}, L.~R., and {Ullom}, J.~N., ``{Microwave
  SQUID multiplexer demonstration for cosmic microwave background imagers},''
  {\em Applied Physics Letters}~{\bf 111},  243510 (Dec. 2017).

\bibitem{Dicker2014}
{Dicker}, S.~R., {Ade}, P.~A.~R., {Aguirre}, J., {Brevik}, J.~A., {Cho}, H.~M.,
  {Datta}, R., {Devlin}, M.~J., {Dober}, B., {Egan}, D., {Ford}, J., {Ford},
  P., {Hilton}, G., {Irwin}, K.~D., {Mason}, B.~S., {Marganian}, P., {Mello},
  M., {McMahon}, J.~J., {Mroczkowski}, T., {Rosenman}, M., {Tucker}, C.,
  {Vale}, L., {White}, S., {Whitehead}, M., and {Young}, A.~H., ``{MUSTANG 2: A
  Large Focal Plane Array for the 100 m Green Bank Telescope},'' {\em Journal
  of Low Temperature Physics}~{\bf 176},  808--814 (Sept. 2014).

\bibitem{Simon2018}
Simon, S.~M., Golec, J.~E., Ali, A., Austermann, J., Beall, J.~A., Bruno, S.
  M.~M., Choi, S.~K., Crowley, K.~T., Dicker, S., Dober, B., Duff, S.~M.,
  Healy, E., Hill, C.~A., Ho, S.-P.~P., Hubmayr, J., Li, Y., Lungu, M.,
  McMahon, J., Orlowski-Scherer, J., Salatino, M., Staggs, S., Wollack, E.~J.,
  Xu, Z., and Zhu, N., ``Feedhorn development and scalability for simons
  observatory and beyond,'' {\em Proc.SPIE}~{\bf 10708},  10708 -- 10708 -- 12
  (2018).

\bibitem{Beckman2018}
Beckman, S. et~al., ``Development of antenna-coupled hemispherical lens arrays
  for the simons observatory,'' in [{\em Millimeter, Submillimeter, and
  Far-Infrared Detectors and Instrumentation for Astronomy
  IX}{\nolinebreak\hspace{0.1em}]},  {\em Proc. SPIE} ,  10708 -- 89 (2018).

\bibitem{Crowley2018}
Crowley, K.~T., Simon, S.~M., Silva-Feaver, M., Goeckner-Wald, N., Ali, A.,
  Austermann, J., Brown, M.~L., Chinone, Y., Cukierman, A., Dober, B., Duff,
  S.~M., Dunkley, J., Errard, J., Fabbian, G., Gallardo, P.~A., Ho, S.-P.~P.,
  Hubmayr, J., Keating, B., Kusaka, A., McCallum, N., McMahon, J., Nati, F.,
  Niemack, M.~D., Puglisi, G., Rao, M.~S., Reichardt, C.~L., Salatino, M.,
  Siritanasak, P., Staggs, S., Suzuki, A., Teply, G., Thomas, D.~B., Ullom,
  J.~N., Vergès, C., Vissers, M.~R., Westbrook, B., Wollack, E.~J., Xu, Z.,
  and Zhu, N., ``Studies of systematic uncertainties for simons observatory:
  detector array effects,'' {\em Proc.SPIE}~{\bf 10708},  10708 -- 10708 -- 27
  (2018).

\bibitem{Henderson2018}
{Henderson}, S. et~al., ``{High mux readout using SLAC electronics for CMB and
  submm },'' in [{\em Millimeter, Submillimeter, and Far-Infrared Detectors and
  Instrumentation for Astronomy IX}{\nolinebreak\hspace{0.1em}]},  {\em
  \procspie},  10708--43 (2018).

\bibitem{Ho2018}
Ho, S.-P. et~al., ``{The universal focal plane module for the Simons
  Observatory},'' in [{\em Millimeter, Submillimeter, and Far-Infrared
  Detectors and Instrumentation for Astronomy IX}{\nolinebreak\hspace{0.1em}]},
   {\em Proc. SPIE} ,  10708 -- 126 (2018).

\bibitem{Parshley2018_2}
Parshley, S.~C. et~al., ``{CCAT-prime:} a novel telescope for sub-millimeter
  astronomy,'' in [{\em Ground-based and Airborne Telescopes
  VII}{\nolinebreak\hspace{0.1em}]},  {\em Proc. SPIE} ,  10700 -- 220 (2018).

\bibitem{Datta2013}
Datta, R., Munson, C.~D., Niemack, M.~D., McMahon, J.~J., Britton, J., Wollack,
  E.~J., Beall, J., Devlin, M.~J., Fowler, J., Gallardo, P., Hubmayr, J.,
  Irwin, K., Newburgh, L., Nibarger, J.~P., Page, L., Quijada, M.~A., Schmitt,
  B.~L., Staggs, S.~T., Thornton, R., and Zhang, L., ``Large-aperture
  wide-bandwidth antireflection-coated silicon lenses for millimeter
  wavelengths,'' {\em Appl. Opt.}~{\bf 52},  8747--8758 (Dec 2013).

\bibitem{Dicker2018}
{Dicker}, S.~R., {Gallardo}, P.~A., {Gundmudsson}, J., {Mauskopf}, P.~D.,
  {Niemack}, M.~D., {Ali}, A., {Ashton}, P.~C., {Coppi}, G., {Devlin}, M.~J.,
  {Galitzki}, N., P., H.~S., {Hill}, C.~A., {Hubmayr}, J., {Keating}, B.,
  {Lee}, A.~T., {Limon}, M., {Matsuda}, F., {McMahon}, J., {Orlowski-Scherer},
  J.~L., {Piccirillo}, L., M., S.~S., {Thornton}, R., {Ullom}, J.~N.,
  {Vavagiakis}, E.~M., {Wollack}, E.~J., {Xu}, Z., and {Zhu}, N., ``{Cold
  optical design for the Large Aperture Simons Observatory telescope},'' in
  [{\em Ground-based and Airborne Telescopes VII}{\nolinebreak\hspace{0.1em}]},
   {\em \procspie},  10700--122 (2018).

\bibitem{Gallardo2018}
{Gallardo}, p., {Matsuda}, F., {Gudmundsson}, J., {Simon}, S., {Koopman},
  B.~J., et~al., ``{Studies of Systematic Uncertainties for Simons Observatory:
  Optical Effects and Sensitivity Considerations},'' in [{\em Millimeter,
  Submillimeter, and Far-Infrared Detectors and Instrumentation for Astronomy
  IX}{\nolinebreak\hspace{0.1em}]},  {\em \procspie},  10708--133 (2018).

\bibitem{Zhu2018}
Zhu, N., Orlowski-Scherer, J.~L., Xu, Z., Ali, A., Arnold, K.~S., Ashton,
  P.~C., Coppi, G., Devlin, M.~J., Dicker, S., Galitzki, N., Gallardo, P.~A.,
  Henderson, S.~W., Ho, S.-P.~P., Hubmayr, J., Keating, B., Lee, A.~T., Limon,
  M., Lungu, M., Mauskopf, P.~D., May, A.~J., McMahon, J., Niemack, M.~D.,
  Piccirillo, L., Puglisi, G., Rao, M.~S., Salatino, M., Silva-Feaver, M.,
  Simon, S.~M., Staggs, S., Thornton, R., Ullom, J.~N., Vavagiakis, E.~M.,
  Westbrook, B., and Wollack, E.~J., ``Simons observatory large aperture
  telescope receiver design overview,'' {\em Proc.SPIE}~{\bf 10708},  10708 --
  10708 -- 15 (2018).

\bibitem{Orlowski-Scherer2018}
Orlowski-Scherer, J.~L., Zhu, N., Xu, Z., Ali, A., Arnold, K.~S., Ashton,
  P.~C., Coppi, G., Devlin, M., Dicker, S., Galitzki, N., Gallardo, P.~A.,
  Keating, B., Lee, A.~T., Limon, M., Lungu, M., May, A., McMahon, J., Niemack,
  M.~D., Piccirillo, L., Puglisi, G., Salatino, M., Silva-Feaver, M., Simon,
  S.~M., Thornton, R., and Vavagiakis, E.~M., ``Simons observatory large
  aperture receiver simulation overview,'' {\em Proc.SPIE}~{\bf 10708},  10708
  -- 10708 -- 14 (2018).

\bibitem{Coppi2018}
Coppi, G., Xu, Z., Ali, A., Galitzki, N., Gallardo, P.~A., May, A.~J.,
  Orlowski-Scherer, J.~L., Zhu, N., Devlin, M.~J., Dicker, S., Keating, B.,
  Limon, M., Lungu, M., McMahon, J., Niemack, M.~D., Piccirillo, L., Puglisi,
  G., Salatino, M., Simon, S.~M., Teply, G., Thornton, R., and Vavagiakis,
  E.~M., ``Cooldown strategies and transient thermal simulations for the
  {Simons Observatory},'' {\em Proc.SPIE}~{\bf 10708},  10708 -- 10708 -- 13
  (2018).

\bibitem{Vavagiakis2018}
{Vavagiakis}, E. et~al., ``{Prime-Cam: A first-light instrument for the
  CCAT-prime telescope},'' in [{\em Millimeter, Submillimeter, and Far-Infrared
  Detectors and Instrumentation for Astronomy IX}{\nolinebreak\hspace{0.1em}]},
   {\em \procspie},  10708--64 (2018).

\bibitem{Wickersheimer1976}
Wickersheimer, D.~J., ``{The Vierendeel},'' {\em Journal of the Society of
  Architectural Historians}~{\bf 35},  54--60 (Mar. 1976).

\bibitem{Inoue2016}
{Inoue}, Y., {Ade}, P., {Akiba}, Y., {Aleman}, C., {Arnold}, K., {Baccigalupi},
  C., {Barch}, B., {Barron}, D., {Bender}, A., {Boettger}, D., {Borrill}, J.,
  {Chapman}, S., {Chinone}, Y., {Cukierman}, A., {de Haan}, T., {Dobbs}, M.~A.,
  {Ducout}, A., {D{\"u}nner}, R., {Elleflot}, T., {Errard}, J., {Fabbian}, G.,
  {Feeney}, S., {Feng}, C., {Fuller}, G., {Gilbert}, A.~J., {Goeckner-Wald},
  N., {Groh}, J., {Hall}, G., {Halverson}, N., {Hamada}, T., {Hasegawa}, M.,
  {Hattori}, K., {Hazumi}, M., {Hill}, C., {Holzapfel}, W.~L., {Hori}, Y.,
  {Howe}, L., {Irie}, F., {Jaehnig}, G., {Jaffe}, A., {Jeong}, O., {Katayama},
  N., {Kaufman}, J.~P., {Kazemzadeh}, K., {Keating}, B.~G., {Kermish}, Z.,
  {Keskitalo}, R., {Kisner}, T.~S., {Kusaka}, A., {Le Jeune}, M., {Lee}, A.~T.,
  {Leon}, D., {Linder}, E.~V., {Lowry}, L., {Matsuda}, F., {Matsumura}, T.,
  {Miller}, N., {Mizukami}, K., {Montgomery}, J., {Navaroli}, M., {Nishino},
  H., {Paar}, H., {Peloton}, J., {Poletti}, D., {Puglisi}, G., {Raum}, C.~R.,
  {Rebeiz}, G.~M., {Reichardt}, C.~L., {Richards}, P.~L., {Ross}, C.,
  {Rotermund}, K.~M., {Segawa}, Y., {Sherwin}, B.~D., {Shirley}, I.,
  {Siritanasak}, P., {Stebor}, N., {Stompor}, R., {Suzuki}, J., {Suzuki}, A.,
  {Tajima}, O., {Takada}, S., {Takatori}, S., {Teply}, G.~P., {Tikhomirov}, A.,
  {Tomaru}, T., {Whitehorn}, N., {Zahn}, A., and {Zahn}, O., ``{POLARBEAR-2: an
  instrument for CMB polarization measurements},'' in [{\em Millimeter,
  Submillimeter, and Far-Infrared Detectors and Instrumentation for Astronomy
  VIII}{\nolinebreak\hspace{0.1em}]},  {\em \procspie} {\bf 9914},  99141I
  (July 2016).

\bibitem{Gudmundsson2015}
{Gudmundsson}, J.~E., {Ade}, P.~A.~R., {Amiri}, M., {Benton}, S.~J., {Bock},
  J.~J., {Bond}, J.~R., {Bryan}, S.~A., {Chiang}, H.~C., {Contaldi}, C.~R.,
  {Crill}, B.~P., {Dore}, O., {Filippini}, J.~P., {Fraisse}, A.~A., {Gambrel},
  A., {Gandilo}, N.~N., {Hasselfield}, M., {Halpern}, M., {Hilton}, G.,
  {Holmes}, W., {Hristov}, V.~V., {Irwin}, K.~D., {Jones}, W.~C., {Kermish},
  Z., {MacTavish}, C.~J., {Mason}, P.~V., {Megerian}, K., {Moncelsi}, L.,
  {Montroy}, T.~E., {Morford}, T.~A., {Nagy}, J.~M., {Netterfield}, C.~B.,
  {Rahlin}, A.~S., {Reintsema}, C.~D., {Ruhl}, J.~E., {Runyan}, M.~C.,
  {Shariff}, J.~A., {Soler}, J.~D., {Trangsrud}, A., {Tucker}, C., {Tucker},
  R.~S., {Turner}, A.~D., {Wiebe}, D.~V., {Young}, E., {Spider Collaboration},
  {Abe}, P.~A.~R., {Amiri}, M., {Benton}, S.~J., {Bock}, J.~J., {Bond}, J.~R.,
  {Bryan}, S.~A., {Chiang}, H.~C., {Contaldi}, C.~R., {Crill}, B.~P., {Dore},
  O., {Filippini}, J.~P., {Fraisse}, A.~A., {Gambrel}, A., {Gandilo}, N.~N.,
  {Hasselfield}, M., {Halpern}, M., {Hilton}, G., {Holmes}, W., {Hristov},
  V.~V., {Irwin}, K.~D., {Jones}, W.~C., {Kermish}, Z., {MacTavish}, C.~J.,
  {Mason}, P.~V., {Megerian}, K., {Moncelsi}, L., {Montroy}, T.~E., {Morford},
  T.~A., {Nagy}, J.~M., {Netterfield}, C.~B., {Rahlin}, A.~S., {Reintsema},
  C.~D., {Ruhl}, J.~E., {Runyan}, M.~C., {Shariff}, J.~A., {Soler}, J.~D.,
  {Trangsrud}, A., {Tucker}, C., {Tucker}, R.~S., {Turner}, A.~D., {Wiebe},
  D.~V., and {Young}, E., ``{The thermal design, characterization, and
  performance of the SPIDER long-duration balloon cryostat},'' {\em
  Cryogenics}~{\bf 72},  65--76 (Dec. 2015).

\bibitem{Hill2018_2}
{Hill}, C.~A., {Kusaka}, A., {Barton}, P., {Bixler}, B., {Droster}, A.~G.,
  {Flament}, M., {Ganjam}, S., {Jadbabaie}, A., {Jeong}, O., {Lee}, A.~T.,
  {Madurowicz}, A., {Matsuda}, F.~T., {Matsumura}, T., {Rutkowski}, A.,
  {Sakurai}, Y., {Sponseller}, D.~R., {Suzuki}, A., and {Tat}, R., ``{A
  Large-Diameter Cryogenic Rotation Stage for Half-Wave Plate Polarization
  Modulation on the POLARBEAR-2 Experiment},'' {\em Journal of Low Temperature
  Physics}  (May 2018).

\bibitem{Kusaka2018}
{Kusaka}, A., {Appel}, J., {Essinger-Hileman}, T., {Beall}, J.~A., {Campusano},
  L.~E., {Cho}, H.-M., {Choi}, S.~K., {Crowley}, K., {Fowler}, J.~W.,
  {Gallardo}, P., {Hasselfield}, M., {Hilton}, G., {Ho}, S.-P.~P., {Irwin}, K.,
  {Jarosik}, N., {Niemack}, M.~D., {Nixon}, G.~W., {Nolta}, M., {Page}, Jr,
  L.~A., {Palma}, G.~A., {Parker}, L., {Raghunathan}, S., {Reintsema}, C.~D.,
  {Sievers}, J., {Simon}, S.~M., {Staggs}, S.~T., {Visnjic}, K., and {Yoon},
  K.-W., ``{Results from the Atacama B-mode Search (ABS) Experiment},'' {\em
  ArXiv e-prints}  (Jan. 2018).

\bibitem{Salatino2018}
{Salatino}, M., {Lashner}, J., {Gerbino}, M., {Simon}, M.~S., {Didier}, J.,
  {Alim}, A., {Ashton}, P.~C., {Bryane}, S., {Chinoe}, Y., {Crowley}, K.~T.,
  {Fabbian}, G., {Galitzki}, N., {Goeckner-Wald}, N., {Gudmundsson}, J.,
  {Hill}, C.~A., {Keating}, B., {Lee}, A.~T., {McMahon}, J., {Miller}, A.~D.,
  {Puglisi}, G., {Reichardt}, C.~L., {Teply}, G., {Xu}, Z., {Zhu}, N., et~al.,
  ``{Studies of Systematic Uncertainties for Simons Observatory: Polarization
  Modulator Related Effects},'' in [{\em Millimeter, Submillimeter, and
  Far-Infrared Detectors and Instrumentation for Astronomy
  IX}{\nolinebreak\hspace{0.1em}]},  {\em \procspie},  10708--143 (2018).

\bibitem{Ade2006}
Ade, P. A.~R., Pisano, G., Tucker, C., and Weaver, S., ``A review of metal mesh
  filters,'' {\em Proc.SPIE}~{\bf 6275},  6275 -- 15 (2006).

\bibitem{S4sci}
{Abazajian}, K.~N., {Adshead}, P., {Ahmed}, Z., {Allen}, S.~W., {Alonso}, D.,
  {Arnold}, K.~S., {Baccigalupi}, C., {Bartlett}, J.~G., {Battaglia}, N.,
  {Benson}, B.~A., {Bischoff}, C.~A., {Borrill}, J., {Buza}, V., {Calabrese},
  E., {Caldwell}, R., {Carlstrom}, J.~E., {Chang}, C.~L., {Crawford}, T.~M.,
  {Cyr-Racine}, F.-Y., {De Bernardis}, F., {de Haan}, T., {di Serego
  Alighieri}, S., {Dunkley}, J., {Dvorkin}, C., {Errard}, J., {Fabbian}, G.,
  {Feeney}, S., {Ferraro}, S., {Filippini}, J.~P., {Flauger}, R., {Fuller},
  G.~M., {Gluscevic}, V., {Green}, D., {Grin}, D., {Grohs}, E., {Henning},
  J.~W., {Hill}, J.~C., {Hlozek}, R., {Holder}, G., {Holzapfel}, W., {Hu}, W.,
  {Huffenberger}, K.~M., {Keskitalo}, R., {Knox}, L., {Kosowsky}, A., {Kovac},
  J., {Kovetz}, E.~D., {Kuo}, C.-L., {Kusaka}, A., {Le Jeune}, M., {Lee},
  A.~T., {Lilley}, M., {Loverde}, M., {Madhavacheril}, M.~S., {Mantz}, A.,
  {Marsh}, D. J.~E., {McMahon}, J., {Meerburg}, P.~D., {Meyers}, J., {Miller},
  A.~D., {Munoz}, J.~B., {Nguyen}, H.~N., {Niemack}, M.~D., {Peloso}, M.,
  {Peloton}, J., {Pogosian}, L., {Pryke}, C., {Raveri}, M., {Reichardt}, C.~L.,
  {Rocha}, G., {Rotti}, A., {Schaan}, E., {Schmittfull}, M.~M., {Scott}, D.,
  {Sehgal}, N., {Shandera}, S., {Sherwin}, B.~D., {Smith}, T.~L., {Sorbo}, L.,
  {Starkman}, G.~D., {Story}, K.~T., {van Engelen}, A., {Vieira}, J.~D.,
  {Watson}, S., {Whitehorn}, N., and {Kimmy Wu}, W.~L., ``{CMB-S4 Science Book,
  First Edition},'' {\em ArXiv e-prints} ,  arXiv:1610.02743 (Oct. 2016).

\bibitem{S4tech}
{Abitbol}, M.~H., {Ahmed}, Z., {Barron}, D., {Basu Thakur}, R., {Bender},
  A.~N., {Benson}, B.~A., {Bischoff}, C.~A., {Bryan}, S.~A., {Carlstrom},
  J.~E., {Chang}, C.~L., {Chuss}, D.~T., {Crowley}, K.~T., {Cukierman}, A., {de
  Haan}, T., {Dobbs}, M., {Essinger-Hileman}, T., {Filippini}, J.~P., {Ganga},
  K., {Gudmundsson}, J.~E., {Halverson}, N.~W., {Hanany}, S., {Henderson},
  S.~W., {Hill}, C.~A., {Ho}, S.-P.~P., {Hubmayr}, J., {Irwin}, K., {Jeong},
  O., {Johnson}, B.~R., {Kernasovskiy}, S.~A., {Kovac}, J.~M., {Kusaka}, A.,
  {Lee}, A.~T., {Maria}, S., {Mauskopf}, P., {McMahon}, J.~J., {Moncelsi}, L.,
  {Nadolski}, A.~W., {Nagy}, J.~M., {Niemack}, M.~D., {O'Brient}, R.~C.,
  {Padin}, S., {Parshley}, S.~C., {Pryke}, C., {Roe}, N.~A., {Rostem}, K.,
  {Ruhl}, J., {Simon}, S.~M., {Staggs}, S.~T., {Suzuki}, A., {Switzer}, E.~R.,
  {Tajima}, O., {Thompson}, K.~L., {Timbie}, P., {Tucker}, G.~S., {Vieira},
  J.~D., {Vieregg}, A.~G., {Westbrook}, B., {Wollack}, E.~J., {Yoon}, K.~W.,
  {Young}, K.~S., and {Young}, E.~Y., ``{CMB-S4 Technology Book, First
  Edition},'' {\em ArXiv e-prints} ,  arXiv:1706.02464 (June 2017).

\end{thebibliography}
\bibliographystyle{spiebib}   

\end{document}